\documentclass[%
 reprint,
 amsmath,amssymb,
 aps,
]{revtex4-2}

\usepackage{graphicx}
\usepackage{dcolumn}
\usepackage{bm}

\begin{document}

\preprint{APS/123-QED}

\title{\boldmath Study of Heavy Hadron Production in Au + Au Collisions at a Center-of-Mass Energy of $\sqrt{s_{NN}}=200$ GeV}%

\author{Sunidhi Saxena}
\affiliation{Department of Physics, Institute of Science, Banaras Hindu University (BHU), \\Varanasi 221005, India}
\author{Rajiv Gupta}%
\affiliation{Department of Physics, Institute of Science, Banaras Hindu University (BHU), \\Varanasi 221005, India}

\author{Gauri Devi}
 \affiliation{Department of Physics, Institute of Science, Banaras Hindu University (BHU), \\Varanasi 221005, India}
\author{Ajay Kumar}
\email{ajay.phy@bhu.ac.in}
\affiliation{Department of Physics, Institute of Science, Banaras Hindu University (BHU), \\Varanasi 221005, India}

\date{\today}

\begin{abstract}
Using the Monte Carlo HYDJET++ model, the transverse momentum ($p_{T}$) spectra of heavy hadrons ($D^{0}$, $\overline{D}^{0}$, $D^{+}$, $D^{-}$ and $\Lambda_{c}$), as well as the nuclear modification factors of $D^{0}$ and $D^{\pm}$, produced in Au + Au collisions at $\sqrt{s_{NN}} = 200$ GeV RHIC energy across various centrality bins, are presented. This study is motivated by the need to understand the centrality dependence of the charm enhancement factor ($\gamma_{c}$) and the roles of different hadronization mechanisms such as coalescence and fragmentation, in charm hadron production. To achieve the best description of heavy hadron production, several input parameters in both the soft and hard components of the model are tuned. The study finds a decreasing trend of $\gamma_{c}$ from central to peripheral collisions and a mass dependence across charm hadrons. Moreover, the model effectively reproduces experimental data of $p_{T}$ spectra at low and intermediate $p_{T}$, capturing key features of charm hadron production in the quark-gluon plasma medium. However, it overpredicts the data at high $p_{T}$, indicating the need for improvements in modeling heavy quark energy loss mechanisms. Further, the nuclear modification factors ($R_{AA}$ and $R_{CP}$) for $D^{0}$ mesons exhibit significant suppression in central collisions, which matches with experimental observations. This highlights the roles of collisional and radiative energy loss due to collective effects, such as coalescence and radial flow. The antiparticle-to-particle and mixed particle ratios are also presented, showing good agreement with experimental data and revealing limitations in baryon production due to the absence of heavy quark coalescence in HYDJET++. Additionally, to demonstrate the performance of the HYDJET++ model relative to other theoretical approaches, its results are compared with the TAMU, SUBATECH, Torino, Duke, and AMPT models. 
\end{abstract}

\maketitle


\section{\label{sec:intro}Introduction}

The curiosity to deeply understand the strong interaction and the formation of matter in the earliest stage of the universe requires the study of Quark Gluon Plasma (QGP) \cite{Busza:2018rrf}. QGP is the deconfined, hot, and dense color-conducting state of matter, which is formed just a few microseconds after the Big Bang. Thus, the study of QGP is an important tool for understanding the properties of nuclear matter under extreme conditions.
In the laboratory, QGP matter is created at high temperatures and nearly zero baryon chemical potential in ultrarelativistic heavy-ion collisions. As the temperature decreases, the QGP phase rapidly transitions into the hadron phase, where new particles are produced due to hadronic interactions. On further decrease of temperature, inelastic interactions between hadrons cease, fixing particle abundances and types; this particular temperature is known as the chemical freeze-out temperature ($T_{ch}$). At a later stage, elastic interactions also cease, marking the kinetic (thermal) freeze-out temperature ($T_{th}$).
Since QGP is formed for a very short time, its properties cannot be studied directly; they are inferred from the particles emerging from the QGP after interacting with the medium. These particles include light hadrons, dileptons, photons, and heavy hadrons.

 Heavy quarks play a vital role in studying QGP. They (charm and beauty quarks) are produced during the initial hard scattering of heavy ions due to their large masses ($m_{b} \approx 4.7$ GeV, $m_{c} \approx 1.3$ GeV \cite{ParticleDataGroup:2014cgo}) \cite{Lin:1994xma, Cacciari:2005rk} and propagate through the medium throughout the entire evolution of the QGP. Heavy quarks gain their mass primarily through coupling with the Higgs field in the electroweak sector, which allows them to remain massive even within the QGP, where chiral symmetry is restored.  In contrast, the masses of light quarks (u, d, s) arise predominantly from the spontaneous breaking of chiral symmetry in Quantum Chromodynamics (QCD)  \cite{Zhu:2006er}. The dynamically generated mass of light quarks vanishes on chirality restoration, which may affect their ability to retain information about the QGP formation. Thus, heavy quarks are valuable probes for investigating the pre-equilibrium phase and transport properties of highly dense nuclear matter through perturbative calculations of heavy quark production cross-sections \cite{Prino:2016cni}.

Some important observables for studying the properties of QGP are the transverse momentum ($p_{T}$) spectra, nuclear modification factors ($R_{AA}$ and $R_{CP}$), the antiparticle-to-particle ratio, and mixed particle ratios. The slope of the $p_{T}$ spectra \cite{Heinz:2013th} at low $p_{T}$ is sensitive to the kinetic freeze-out temperature and radial collective flow, providing insights into its thermal and hydrodynamic behavior. At high $p_{T}$, heavy quarks interact with the highly excited particles of the QGP medium; thereby losing their energy and momentum, creating a wake. This phenomenon is known as ``jet quenching'', where jets are collimated sprays of highly excited hadrons. The energy loss occurs either through interactions with the medium constituents or via gluon radiation \cite{Gyulassy:1993hr}. This energy loss is observed through nuclear modification factors ($R_{AA}$ and $R_{CP}$). The nuclear modification factor ($R_{AA}$) is defined as the ratio of the yield of heavy-flavored hadrons in Au + Au collisions to that in p + p collisions (where QGP is not formed), scaled by the mean number of binary nucleon-nucleon sub-collisions. Similarly, $R_{CP}$ represents the ratio of heavy hadron yields between central and peripheral Au + Au collisions, normalized by the number of binary collisions. These factors were first measured at the Relativistic Heavy Ion Collider (RHIC) and the Large Hadron Collider (LHC) during central nucleus-nucleus collisions \cite{PHENIX:2001hpc, ALICE:2012sxy}.

Charm and bottom quarks experience different energy losses when colliding with the QGP medium \cite{Braaten:1991we}. The energy loss of heavy quarks via gluon radiation is smaller than that of light quarks at angles ($\theta$) below the ratio of the quark mass ($M$) to its energy ($E$). This phenomenon is known as the ``dead cone effect'' \cite{Dokshitzer:2001zm}. The suppression factor quantifies this difference and is described as follows \cite{Dokshitzer:2001zm}:

\begin{equation}
    d P_{HQ} = d P_{0}(1+ \frac{\theta_{0}^{2}}{\theta^{2}})^{-2}, \theta_{0} \equiv \frac{M}{E}
\end{equation}
Here, $dP_{HQ}$ represents the distribution of soft gluons radiated by heavy quarks, and $dP_{0}$ denotes the standard bremsstrahlung distribution. As a result, a significantly larger number of heavy quarks are produced compared to light quarks within the QGP medium \cite{Baier:2001yt}.

After the in-medium evolution of heavy quarks, when the QGP medium has cooled down, charm quarks hadronize into various open charm hadrons, such as $D^{0}$, $D^{\pm}$, $D_{s}^{\pm}$, and $\Lambda_{c}^{\pm}$. The hadronization process incorporates two primary mechanisms such as coalescence and fragmentation. The coalescence of minijet partons with QGP medium constituents was first described in Ref. \cite{Lin:2002rw}. At $2 < p_{T} < 6 ~\text{GeV/c}$ range, hard partons coalesce with soft partons within the QGP to form hadrons \cite{Greco:2003mm, Greco:2003xt}. Due to the coalescence mechanism, the baryon-to-meson ratio increases for charmed hadrons. A similar trend is observed in light hadrons \cite{STAR:2006uve}. While minijet partons produced in the initial stages of the heavy-ion collision combine with quark-antiquark pairs present in the QGP medium to form high-transverse momentum ($p_{T} > 6 ~\text{GeV/c}$) hadrons. This transformation of heavy quarks into high-$p_{T}$ hadrons is modeled using the fragmentation function \cite{Fries:2003vb}.

Further, the anti-particle-to-particle ratio should be balanced to produce a low baryon density, as expected within the QGP medium. This ratio is useful for gaining insights into the baryon chemical potential ($\mu_{B}$) because the mass term in the particle and anti-particle partition functions is the same \cite{Cleymans:2006xj}. The relation between $\mu_{B}$ and the ratio is given by:
\begin{equation}
    \overline{h}/h \propto \exp{[-2(B\mu_{B} + S\mu_{S})/T]},
\end{equation}
where $B$ is baryon quantum numbers, $S$ is the strange quantum numbers of the particles, $\mu_{S}$ is the strange chemical potential, and $T$ is the chemical freeze-out temperature.

This study is focused on the properties of heavy hadrons except strange-charm hadrons; therefore, this ratio depends on $\mu_{B}$ only. For small values of $\mu_{B}$, the ratio has a negligible dependence on temperature, making the antiparticle-to-particle ratio an ineffective probe for studying $T_{ch}$, while it remains an important observable for $\mu_{B}$ studies.

The mixed particle ratio serves as an effective thermometer for determining the chemical freeze-out temperature \cite{Cleymans:2006xj}. Its relation with temperature is given by:
\begin{equation}
    h_{1}/h_{2} \propto (\frac{m_{h_{1}}}{m_{h_{2}}})^{(3/2)}\exp{[-\frac{m_{h_{1}}-m_{h_{2}}}{T}]}\exp{[-\frac{\mu_{S}}{T}]},
\end{equation}
where $h_{1}$ and $h_{2}$ are the spectra of two hadrons. $m_{h_{1}}$ and $m_{h_{2}}$ are the masses of the two hadrons. The absence of strange particle in this study reduces the second exponential term to unity. Therefore, the chemical freeze-out temperature depends entirely on the masses of the two hadrons.

The production of hadrons containing heavy quarks has been observed in Au + Au collisions at a center-of-mass energy of 200 GeV by the STAR collaboration \cite{Tlusty:2012vr}. It has been observed that $D^{0}$ mesons undergo kinetic freeze-out earlier than the light hadrons (pions and protons), but on a similar timeline as strange hadrons \cite{STAR:2018zdy}. This is because the $D^{0}$ meson, being heavy, interacts less with the medium compared to light hadrons while exhibiting a similar radial flow to strange hadrons. Furthermore, as charm quarks pass through the medium, they lose energy due to strong interactions with the medium at high $p_{T}$. While at low $p_{T}$, the $D^{0}$ meson yield increases due to the collective motion of charm quarks with the medium \cite{STAR:2014wif}. Additionally, $D^{0}$ mesons follow the number-of-constituent-quark scaling \cite{Licenik:2020cjc}.

The production of charmed baryons $\Lambda_{c}$ was first observed in Au + Au collisions at $\sqrt{s_{NN}} = 200 ~\text{GeV}$ at mid-rapidity \cite{STAR:2019ank}. The ratio of charmed baryons to $D^{0}$ mesons is greater than the predictions from PYTHIA simulations for p + p collisions, since QGP is absent in p + p collision \cite{Zhou:2017ikn}. Additionally, charm-strange mesons were first observed at mid-rapidity in Au + Au collisions at 200 GeV, resulting from the coalescence of charm quarks with strange quarks \cite{STAR:2021tte}. Other hadrons from the total charm hadron production are $D^{\pm}$ mesons, which are found to follow a production mechanism similar to that of $D^{0}$ mesons \cite{Vanek:2022ekr}.

Various theoretical models have been proposed to explain the experimental results. These models are TAMU \cite{He:2011qa}, SUBATECH \cite{Gossiaux:2010yx}, Torino \cite{Alberico:2011zy}, Duke \cite{Cao:2013ita}, LANL \cite{Sharma:2009hn}, AMPT \cite{Lin:2004en}, and HYDJET++ \cite{Lokhtin:2008xi}. The TAMU model \cite{He:2011qa} uses non-perturbative T-matrix methods \cite{Riek:2010fk, Riek:2010py} and Fokker-Planck-Langevin dynamics to describe heavy quark transport in a hydrodynamic QGP medium. This model explains the ``flow bump'' in the $R_{AA}$ of $D^{0}$ mesons but does not account for the radiative energy loss of heavy quarks at high $p_{T}$ and hadronic rescattering.
 SUBATECH \cite{Gossiaux:2010yx} incorporates Hard Thermal Loop (HTL)-based calculation for charm-medium interactions with collisional and radiative energy loss. This model reproduces $R_{AA}$ and $v_{2}$, but does not clarify the individual contributions of collisional and radiative energy loss. The Torino group \cite{Alberico:2011zy} combines POWHEG-based initial production \cite{Frixione:2007nu, Frixione:2007nw} with HTL+pQCD transport \cite{Combridge:1978kx} and Langevin evolution \cite{Beraudo:2009pe} but omits coalescence. This model reproduces high-$p_{T}$ ($p_{T} \gtrsim 3~\text{GeV/c}$) $R_{AA}$ results, but does not describe low-$p_{T}$ ($p_{T} \lesssim 3~\text{GeV/c}$) and peripheral collision results. The Duke group \cite{Cao:2013ita} employs leading-order pQCD \cite{Owens:1986mp} with shadowing (EPS08 parameterization) \cite{Eskola:2009yy}, Langevin dynamics augmented with a recoil force term to include the effects of gluon radiation via higher-twist energy loss calculations \cite{Guo:2000nz, Majumder:2009ge, Zhang:2003wk, Qin:2012fua}, and a fragmentation-plus-recombination hadronization scheme \cite{Lin:2003jy, Greco:2003vf, Oh:2009zj, Han:2012hp} using viscous hydrodynamics \cite{Qiu:2011hf}. This model highlights the role of medium-induced gluon radiation in the $R_{AA}$ of heavy quarks at high $p_{T}$ and shows how nuclear shadowing affects $R_{AA}$ across the entire $p_{T}$ range. The LANL group \cite{Sharma:2009hn} emphasizes meson dissociation and partonic energy loss to explain the suppression of non-photonic electrons. A multi-phase transport (AMPT) model \cite{Lin:2004en} uses HIJING for initial parton production \cite{Gyulassy:1994ew}, ZPC model for partonic interactions \cite{Zhang:1997ej}, and Lund string fragmentation for the fragmentation mechanism of hadronization \cite{Andersson:1997xwk}. Its string melting version (AMPT-SM) \cite{Lin:2001zk} is used for the quark coalescence mechanism of hadronization, with final hadron interactions modeled by the ART hadronic transport model \cite{Li:2001xh}. An improved AMPT version \cite{Lin:2021mdn} includes heavy quark cross-sections and removes the $p_{T}$ cutoff for initial production. This version performs well in small systems like p + p collisions. For larger systems like Au + Au, EPS09 shadowing \cite{Helenius:2012wd} is added. It enhances low $p_{T}$ $D^{0}$ yields but leads to overprediction at low $p_{T}$ and underprediction above $p_{T} =$ 2.5 GeV/c.

Finally, the HYDJET++ model \cite{Lokhtin:2008xi} combines thermal and non-thermal components with  a FAST MC generator \cite{Amelin:2006qe, Amelin:2007ic} and PYQUEN generators \cite{Lokhtin:2005px} respectively, modeling a longitudinally expanding QGP. The HYDJET++ model is described in detail in section \ref{sec:model}. HYDJET++ model has successfully described strange hadron production \cite{Singh:2023bzm, Devi:2023wih, Devi:2024uis}. It also effectively describes the early thermal freeze-out of charmed mesons compared to light hadrons at RHIC energy \cite{Lokhtin:2017rvj}. At LHC energy, $D$ mesons thermalize simultaneously with light hadrons, whereas the $J/\psi$ meson undergoes thermal freeze-out earlier than light hadrons. This behavior is due to the increased interaction cross-section of $D$ mesons at LHC energy. These results are obtained under the assumption that $T_{th} = T_{ch}$ \cite{Lokhtin:2017rvj}. Although the previous study has achieved notable successes, it does not capture the centrality dependence of the charm enhancement factor ($\gamma_{c}$) for heavy hadrons and does not account for the role of different hadronization mechanisms, such as coalescence and fragmentation. In this paper, we discussed these aspects, including the variation of the $\gamma_{c}$ with centrality, the impact of different chemical and thermal freeze-out temperatures, and the role of various hadronization mechanisms. This study will be helpful in defining how the production of charm quarks depends on the size of the QGP fireball and how different hadronization mechanisms are responsible for the production of different charm hadrons. 

The structure of this paper is as follows: Section \ref{sec:model} describes the model framework used to produce heavy quark spectra, starting from the initial production of partons through their evolution and energy loss within the medium, and finally, to hadronization and decay. In Section \ref{sec:result}, we present our results, which include the invariant yields of $D^{0}$, $\overline{D}^{0}$, $D^{\pm}$ mesons, and $\Lambda_{c}$ baryons at mid-rapidity ($|y| < 1$) as a function of transverse momentum ($p_{T}$) for various centrality classes. We also discuss the nuclear modification factors ($R_{AA}$ of $D^{0}$ and $D^{\pm}$ mesons and $R_{CP}$ of $D^{0}$ meson), their variation from central to peripheral collisions, the anti-particle to particle ratio for total charm conservation, and the mixed particle yield ratio as a function of $p_{T}$. All results from the HYDJET++ model are compared with the $D^{0}$, $\overline{D}^{0}$ production \cite{STAR:2018zdy}, $D^{\pm}$ production \cite{Vanek:2022ekr} and $\Lambda_{c}$ baryon production \cite{STAR:2019ank} measured by the STAR experiment in Au + Au collisions at $\sqrt{s_{NN}} = 200~\text{GeV}$. Finally, we summarize the key findings in Section \ref{sec:summ}.

\section{\label{sec:model}The HYDJET++ model framework}

The heavy-ion event generator HYDJET++ (the successor of HYDJET \cite{Lokhtin:2005px}) combines the description of the soft hydro-type state (``thermal'' component) with the hard state resulting from medium-modified multi-parton fragmentation (``non-thermal'' component). In the HYDJET++ model, the soft and hard states are treated independently.

The soft component of the HYDJET++ model represents the ``thermal'' hadronic state produced on the chemical and thermal freeze-out hypersurfaces, derived from a parameterized relativistic hydrodynamic model with predefined freeze-out conditions, using the FASTMC event generator \cite{Amelin:2006qe, Amelin:2007ic}. The statistical hadronization approach \cite{Andronic:2003zv, Andronic:2006ky} is employed to define the thermal production of charm hadrons.

The production of thermal hadrons involves the following steps: First, the four-momentum of hadrons is generated within the rest frame of the quark-gluon liquid using the equilibrium distribution function. Next, the spatial position and four-momentum of the fluid element are generated. This is followed by the generation of the boost for the four-momentum of a hadron within the event’s center-of-mass frame. Resonance decays are taken from the SHARE particle decay table \cite{Torrieri:2004zz}.

The (partially) equilibrated Lorentz-invariant distribution function ($f_{c}$) is defined as \cite{Lokhtin:2016xnl}:
\begin{equation}
    f_{c}(p^{*0};T,\gamma_{c}) = \frac{\gamma_{c}^{n_{c}}g_{i}}{exp(p^{*0}/T) \pm 1},
\end{equation}

Here, $p^{*0}$ is the charmed hadron energy, $T$ is the freeze-out temperature, where the hydrodynamic expansion of the QGP fireball ends, and $\gamma_{c}$ is the charm enhancement factor or charm fugacity ($\gamma_{c} \ge 1$). The parameter $n_{c}$ defines the number of charm quarks and anti-quarks present in a hadron, while $g_{i}$ is the spin degeneracy factor ($g_{i} = 2J_{i} + 1$). The `+' sign in the denominator applies to fermions, whereas the `-' sign applies to bosons. The factor $\gamma_{c}$ is the free parameter of the model. The mean multiplicity of charmed hadrons ($\overline{N_{c}}$) propagating through the QGP medium in Minkowski space is calculated using the thermal volume approximation and is given by the following equation \cite{Lokhtin:2016xnl, Akkelin:2005ms}:
\begin{equation}
    \overline{N_{c}} = \rho_{c}^{eq}(T)V_{eff}, ~~\rho_{c}^{eq}(T) = \int d^{3}p^{*}f_{c}(p^{*0};T,\gamma_{c}), 
\end{equation}
where $\rho_{c}^{eq}$ is the charm hadron number density at the freeze-out temperature $T$, and $V_{eff}$ is the total effective volume from which charm hadrons are emitted. $V_{eff}$ depends on the impact parameter of the heavy-ion collision as \cite{Lokhtin:2016xnl}: 
\begin{widetext}
 \begin{equation}
   V_{eff}  = \tau \int_{0}^{2\pi} d\phi \int_{0}^{R(b,\phi)} \sqrt{1 + \delta(b)tanh^{2}Y_{T}(r,b)cos2\phi} ~ coshY_{T}(r,b)rdr \int_{\eta_{min}}^{\eta_{max}}Y_{L}\eta d\eta ~,
\end{equation}   
\end{widetext}

where $\tau$ is the constant proper time of freeze-out, $R(b,\phi)$ is the transverse radius of the fireball in the $\phi$ direction, $\delta(b)$ defines the momentum anisotropy, and $Y_{T}$ and $Y_{L}$ represent the linear profile of transverse flow rapidity and the Gaussian profile of longitudinal flow rapidity.
In HYDJET++, different chemical freeze-out ($T_{ch}$) and thermal freeze-out ($T_{th}$) temperatures are used. At temperature $T_{ch}$, particle number ratios are fixed, and then the effective volume and momentum spectra of hadrons are computed at temperature $T_{th}$, indicating that $T_{th} < T_{ch}$. 

The hard part of the HYDJET++ model is the same as in the HYDJET model \cite{Lokhtin:2005px, Lokhtin:2007ga}. It is based on the PYTHIA \cite{Sjostrand:2006za} and PYTHIA QUENCHED (PYQUEN) \cite{Lokhtin:2005px} event generators. PYTHIA generates the initial parton spectra in hard nucleon-nucleon (NN) scattering. The PYQUEN partonic energy loss model generates impact parameter-dependent binary nucleonic collision vertices obtained using the Glauber model \cite{Loizides:2017ack}. The impact parameter-dependent mean number of jets ($\overline{N_{AA}^{jet}}$) (which includes heavy quarks) produced in heavy-ion collisions in the presence of shadowing is defined as \cite{Lokhtin:2016xnl}:
\begin{widetext}
    \begin{equation}
\overline{N_{AA}^{jet}}(b,\sqrt{s},p_{T}^{min}) = 
\int_{p_{T}^{min}} dp_{T}^{2} \int dy 
\frac{d\sigma_{NN}^{hard}(p_{T},\sqrt{s})}{dp_{T}^{2}dy} \int_{0}^{2\pi} d\psi \int_{0}^{\infty} r\,dr\, 
T_{A}(r_{1})T_{A}(r_{2}) S(r_{1},r_{2},p_{T},y),
\end{equation}
\end{widetext}

Here, $p_{T}^{min}$ is the minimum transverse momentum transfer in NN collisions. The term $d\sigma_{NN}^{hard}(p_{T},\sqrt{s})/dp_{T}^{2}dy$ represents the differential cross-section of the hard process in NN collision at the center of mass energy ($\sqrt{s}$) of the colliding beams, evaluated at $p_{T}^{min}$. The quantities $r_{1}$ and $r_{2}$ are the transverse distances from the centers of the two nuclei to the jet production vertex, respectively, such that $r_{1}+r_{2} = b$, where $b$ is the impact parameter. $T_{A}$ denotes the nuclear thickness function. $S$ is the shadowing factor, which is further defined as \cite{Lokhtin:2008xi}:
\begin{equation}
    S(r_{1},r_{2},p_{T},y) = S_{A}^{i}(x_{1},Q^{2},r_{1})S_{A}^{j}(x_{2},Q^{2},r_{2}),
\end{equation}
Here, $S_{A}^{i}$ and $S_{A}^{j}$ are the ratios of nuclear to nucleon parton distribution functions (PDFs) for partons i and j, respectively. The variables $x_{1}$ and $x_{2}$ are the momentum fractions of the initial partons from the participating nuclei in the hard scattering. These momentum fractions are defined within the relation $Q^{2}=x_{1}x_{2}s$, where 
\begin{equation}
s=2m_{0}E+m_{0}^{2}+m_{q}^{2},    
\end{equation}  
Here, E and $m_{q}$ are the energy and mass of the heavy quark, respectively, and $m_{0}$ is the effective mass of the thermal parton. The shadowing factor $S\le 1$ corresponds to the nuclear shadowing effect on the PDF.

After that, the re-scattering of partons within the dense QGP medium is simulated using a parton path algorithm, including both collisional and radiative energy loss \cite{Lokhtin:2014vda}. The collisional energy loss is due to elastic scattering among the partons and is calculated only in the high-momentum transfer limit \cite{Lokhtin:2005px, Braaten:1991jj, Lokhtin:2000wm, Lokhtin:2008xi}. The low-momentum scattering is negligible in the total collisional energy loss compared to high-momentum scattering. This energy loss is defined as \cite{Lokhtin:2016xnl}:
\begin{equation}
    \frac{dE^{coll}}{dl} = \frac{1}{4T\lambda\sigma} \int_{\mu_{D}^{2}}^{t_{max}}dt\frac{d\sigma}{dt}t, ~~\frac{d\sigma}{dt}  \cong C \frac{2\pi\alpha_{s}^{2}(t)}{t^{2}}\frac{E^{2}}{E^{2}-m_{q}^{2}},
\end{equation}

where $\lambda=1/\sigma\rho$ is the mean free path within the medium, $\sigma$ is the integral cross section of parton interaction, and $\rho$ is the medium density at temperature T, with $\rho \propto T^{3}$. The Debye screening mass squared ($\mu_{D}^{2}$) is defined as $\mu_{D}^{2}(T)\simeq 4\pi\alpha_{s}T^{2}(1+N_{f}/6)$, where $N_{f}$ represents the number of active quark flavors. The maximum momentum transfer squared ($t_{max}$) is given by the following equation:
\begin{equation}
t_{max} = [s-(m_{q}+m_{0})^{2}][s-(m_{q}-m_{0})^{2}]/s    
\end{equation}

 Here, $\alpha_{s}$ is the QCD running coupling constant, and C=9/4,1,4/9 is the multiplication factor corresponding to gg, gq, and qq scatterings, respectively.

The radiative energy loss is defined by the Baier-Dokshitzer-Mueller-Schiﬀ (BDMS) framework \cite{Baier:1999ds, Baier:2001qw}, with a simple generalization to the heavy quark case based on the ``dead-cone'' approximation \cite{Lokhtin:2016xnl}:
\begin{widetext}
  \begin{equation}
    \frac{dE^{rad}}{dl}\Bigr|_{m_{q}\neq 0} = \frac{1}{(1+ (\beta\omega)^{3/2})^{2}}\frac{dE^{rad}}{dl}\Bigr|_{m_{q}=0}, ~ \beta=(\frac{\lambda}{\mu_{D}^{2}})^{1/3}(\frac{m_{q}}{E})^{4/3},
\end{equation}  
\end{widetext}

where

\begin{equation}
 \frac{dE^{rad}}{dl}\Bigr|_{m_{q}=0} = \frac{2\alpha_{s}\mu_{D}^{2}C_{R}}{\pi L} \int_{\mu_{D}^{2}\lambda_{g}}^{E} d\omega [1-x+\frac{x^{2}}{2}] ln|cos(\omega_{1}\tau
 _{1})|, 
\end{equation}
 
\begin{equation}
 \omega_{1} = \sqrt{i(1-x+\frac{C_{R}}{3}x^{2})\overline{\kappa}ln \frac{16}{\overline{\kappa}}},   
\end{equation}

\begin{equation}
     \overline{\kappa} = \frac{\mu_{D}^{2}\lambda_{g}}{\omega(1-x)},
\end{equation}
where $C_{R} = 4/3$ is the quark color factor, $L$ is the path length, $x$ is the ratio of radiated gluon energy to quark energy, defined as $x=\omega/E$, and $\tau_{1}$ is defined as $\tau_{1}= L/(2\lambda_{g})$, where $\lambda_{g}$ is the gluon mean free path.

The final hadronization is described by the Lund string model \cite{Andersson:1997xwk} for hard partons and in-medium emitted gluons. 

In this study, our goal is to analyze the production of heavy hadrons from soft and hard components. The default input parameters of the HYDJET++ model in the Au + Au collision at $\sqrt{s_{NN}} = 200$ GeV are $T_{ch} = 165$ MeV, $T_{th} = 100$ MeV, ${Y_{l}}^{max} = 3.3$, ${Y_{T}}^{max} = 1.1$ and ${p_{T}}^{min} = 3.55$.  The default model is unable to describe heavy hadron production because heavy hadrons thermalize earlier than light hadrons \cite{STAR:2018zdy}. Therefore, we have specifically tuned the model for heavy hadrons. All the free parameters of the model are optimized to ensure they do not influence the charged hadron spectra \cite{Lokhtin:2015dka, Lokhtin:2016xnl, Devi:2024uis}. The tuned parameters of the HYDJET++ model for the best definition of heavy hadrons are given in table \ref{table0}.

\begin{table}[tbp]
    \centering
    \begin{tabular}{|c|c| c|c|c|c|}
 \hline \hline    Particle type & $T_{ch}$ & $T_{th}$ & ${Y_{l}}^{max}$ & ${Y_{T}}^{max}$ & ${p_{T}}^{min}$ \\
 \hline \hline
                 Heavy Hadrons & 165 MeV  & 125 MeV   & 2.9   & 0.78   & 4.8 GeV/c\\
     \hline
    \end{tabular}
    \caption{\label{table0} Tuned parameters of HYDJET++ for modeling heavy hadron production.}
    
\end{table}

Moreover, we have also adjusted the charm enhancement factor ($\gamma_{c}$). Table \ref{table1} represents the centrality-dependent $\gamma_{c}$ values. It shows that $\gamma_{c}$ decreases from central to peripheral collisions. This is attributed to the reduced system size or $N_{\text{part}}$, resulting in fewer charm quarks being produced in peripheral collisions compared to central collisions. For the $D^{0}$ meson, $\gamma_{c} = 6.5$ and $\Lambda_{c}$ baryon, $\gamma_{c} = 10$ for the most central (0-10\%) collision. Thus, $\gamma_{c}$ also shows mass dependence.

Additionally, in various MC event generators and experiments, the geometrical quantities ($\langle N_{\text{part}}\rangle$ and $\langle N_{\text{coll}}\rangle$) are calculated from the Monte Carlo Glauber model \cite{Loizides:2017ack}. These geometrical quantities are unaffected by tuning input parameters such as $T_{th}$ and $T_{ch}$. This is why $\langle N_{\text{part}}\rangle$ and $\langle N_{\text{coll}}\rangle$ values from the modified HYDJET++ model are compatible with the STAR experiment data \cite{STAR:2018zdy}.

\begin{table}[tbp]
    \centering
    \begin{tabular}{|c|c| c|c|c|c|}
 \hline \hline    $centrality$ & 0-10\% & 10-20\% & 20-40\% & 40-60\% & 60-80\% \\
 \hline \hline
                  $\langle N_{\text{coll}}\rangle$ & 895.8  & 562.6   & 272.2   & 79.49   & 14.98\\
                  $\langle N_{\text{part}}\rangle$ & 317.3  & 223.8   & 129.3   & 51.76   & 14.25\\
                  $\gamma_{c}$ & 6.5    & 5.5     & 5.0     & 3.5     & 3.4\\
                  $\langle p_{T}\rangle$    & 1.364  & 1.358   & 1.353   & 1.352   & 1.336 \\
    
     \hline
    \end{tabular}
    \caption{\label{table1} Average number of collisions ($\langle N_{\text{coll}}\rangle$), average number of participants ($\langle N_{\text{part}}\rangle$), charm enhancement factor ($\gamma_{c}$), and average transverse momentum ($\langle p_{T}\rangle$) from the modified HYDJET++ model.}
    
\end{table}

\section{Results}
\label{sec:result}

\subsection{The transverse momentum ($p_{T}$) spectra}

\begin{figure*}
 \includegraphics[width=15cm]{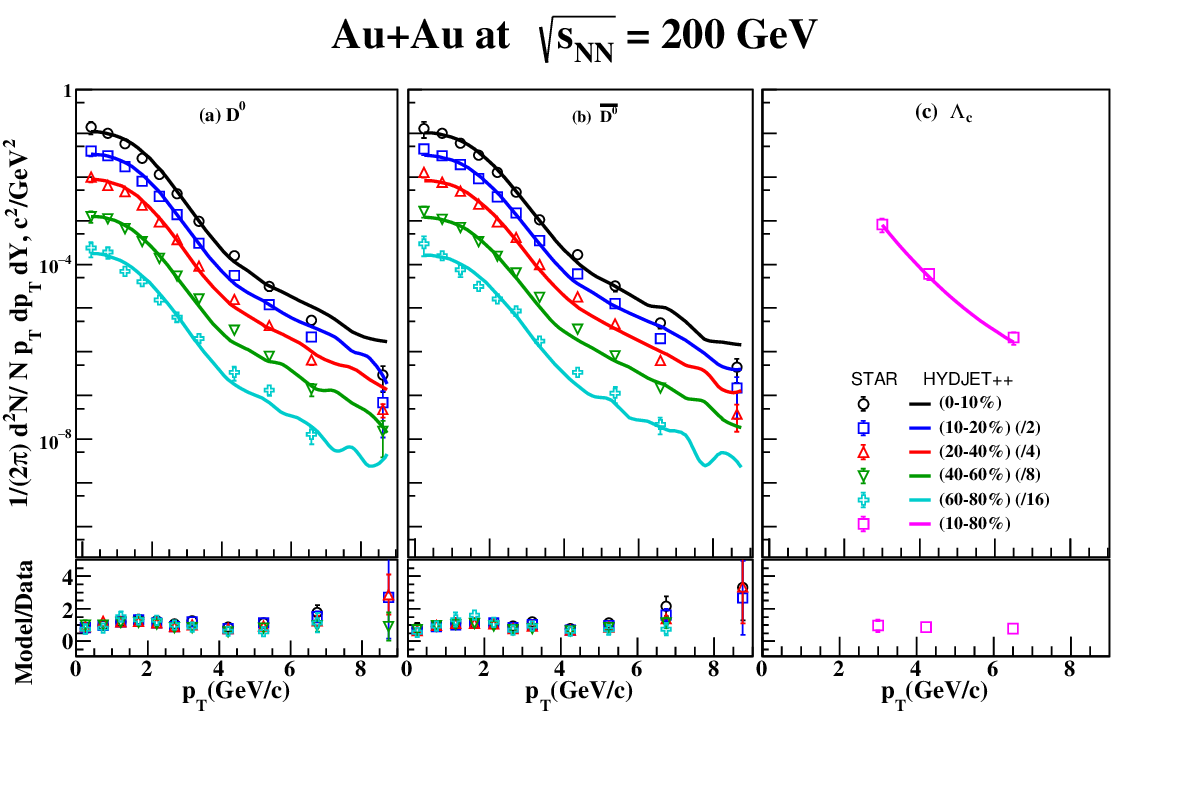}
\caption{\label{figure1}Transverse momentum spectra of $D^{0}$, $\overline{D}^{0}$ mesons, and $\Lambda_{c}$ baryons at mid-rapidity $(\lvert y \rvert < 1)$ for various centrality classes. Open markers represent data from STAR measurements \cite{STAR:2018zdy, STAR:2019ank}, while lines indicate results from the HYDJET++ model.}   
\end{figure*}

\begin{figure*}
 \includegraphics[width=14cm]{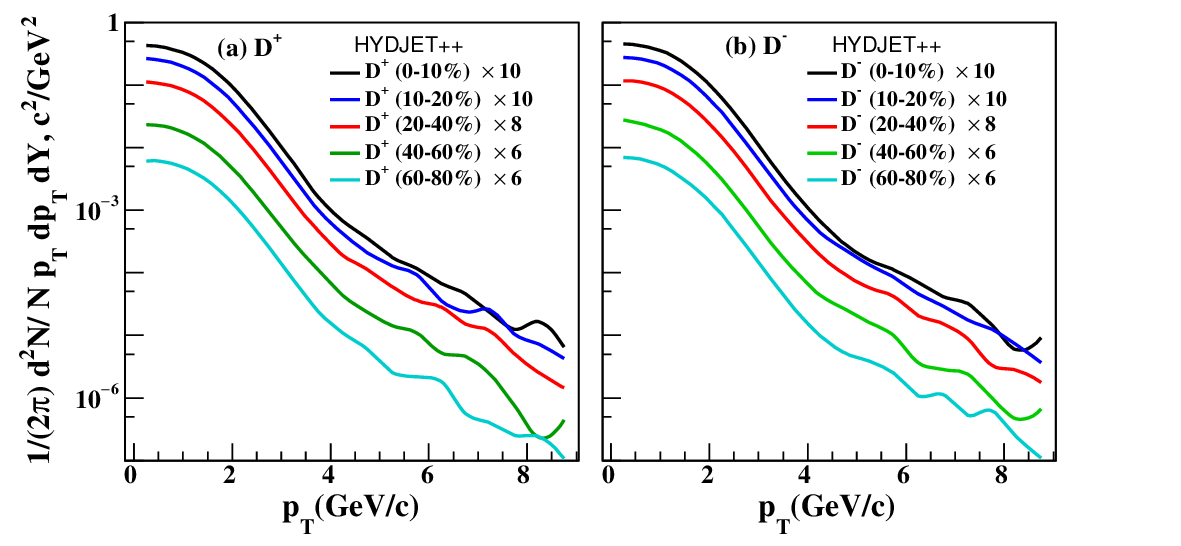}
\caption{\label{figure2}Transverse momentum spectra of (a) $D^{+}$ and (b) $D^{-}$ mesons at mid-rapidity $(\lvert y \rvert < 1)$ for various centrality classes using the HYDJET++ model.}   
\end{figure*}

Figure \ref{figure1} represents the transverse momentum spectra of $D^{0}$ (Figure \ref{figure1}(a)) and $\overline{D}^{0}$ (Figure \ref{figure1}(b)) mesons at mid-rapidity $(\lvert y \rvert < 1)$ in the 0-10\%, 10-20\%, 20-40\%, 40-60\%, and 60-80\% centrality bins. Figure \ref{figure1}(c) represents the transverse momentum spectra of the $\Lambda_{c}$ baryon in the 10-80\% centrality bin. For better visibility, the $p_{T}$ spectra for some centrality bins are scaled by arbitrary factors, as indicated in the legend. The simulated results are compared with experimental data from the STAR experiment \cite{STAR:2018zdy, STAR:2019ank}. The slope of the $p_{T}$ spectra is inversely proportional to the effective temperature of the QGP fireball. As we move from the most peripheral to the most central collisions, the slope of the $p_{T}$ spectra decreases. This indicates that central collisions have a higher QGP temperature than non-central collisions, which is consistent with STAR experimental data \cite{STAR:2018zdy, STAR:2019ank}.

\begin{figure*}
\includegraphics[width=15cm]{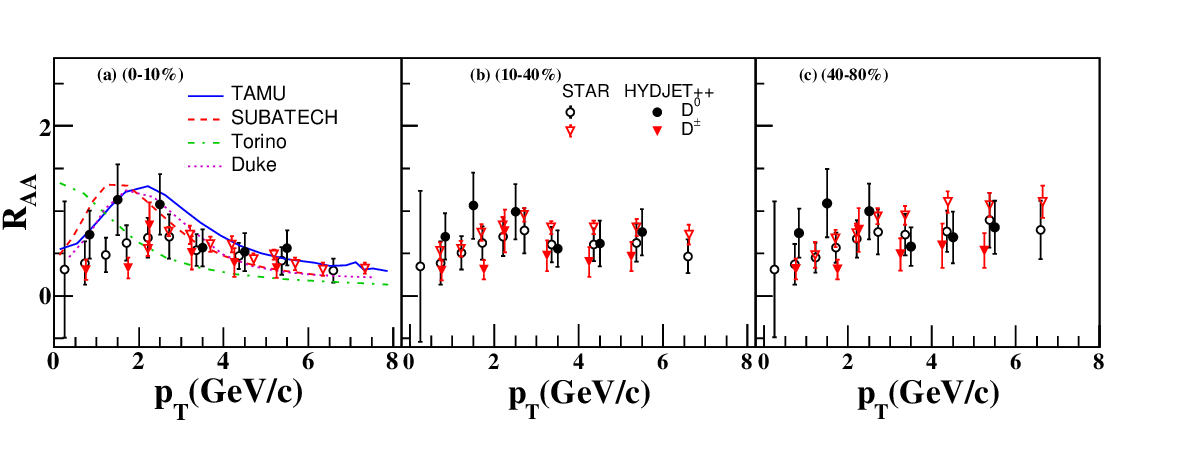}
\caption{\label{figure3} $R_{AA}$ of $D^{0}$ and $D^{\pm}$ mesons as a function of $p_{T}$ for 0-10\%, 10-40\% and 40-80\% centrality classes in Au + Au collisions at $\sqrt{s_{NN}} = 200$ GeV. Open markers represent data from STAR measurements \cite{STAR:2018zdy, Vanek:2022ekr}, while closed markers indicate our model results. Different lines show the results from the TAMU \cite{He:2011qa}, SUBATECH \cite{Gossiaux:2010yx}, Torino \cite{Alberico:2011zy}, and Duke \cite{Cao:2013ita} models for $D^{0}$ meson $R_{AA}$ in the most central collisions.}
\end{figure*}

At low $p_{T}$, the $p_{T}$ spectra follow an exponential distribution and at high $p_{T}$ (typically above ~4 GeV/c), the spectrum transitions from an exponential to a power-law behavior due to contributions from hard scatterings \cite{Gupta:2020naz}.
Additionally, we represent the model-to-data ratio in the lower panel of figure \ref{figure1}. From this ratio, it is evident that the HYDJET++ model describes the $p_{T}$ spectra of $D^{0}$ and $\overline{D}^{0}$ mesons well at low and intermediate $p_{T}$. However, at high $p_{T}$, the model overpredicts the experimental data across all centrality bins. This overprediction shows the need for refinement of energy loss of charm quarks at high $p_{T}$. The $p_{T}$ spectra of $\Lambda_{c}$ baryon also match well with the STAR results \cite{STAR:2019ank}.

Further, figure \ref{figure2} shows the $p_{T}$ spectra of $D^{\pm}$ mesons at mid-rapidity $(\lvert y \rvert < 1)$ in the 0-10\%, 10-20\%, 20-40\%, 40-60\%, and 60-80\% centrality bins in Au + Au collisions at $\sqrt{s_{NN}} = 200 ~ \text{GeV}$. Multiplication factors for each centrality bin are indicated in the legend. Experimental data for $D^{\pm}$ mesons are unavailable; therefore, only the HYDJET++ model results are presented. It is observed that $D^{\pm}$ mesons follow similar centrality dependence as $D^{0}$ mesons, indicating that $D^{\pm}$ mesons thermalize at the same time as $D^{0}$ mesons.

 The average transverse momentum ($\langle p_{T}\rangle$) of heavy hadrons for all centrality bins is listed in table \ref{table1}. It is observed that $\langle p_{T}\rangle$ has minimal dependence on collision centrality. For most central collisions, $\langle p_{T}\rangle$  is larger than in peripheral collisions because high QGP temperature corresponds to a longer lifetime of the fireball, producing a large pressure gradient in an outward direction, which results in sizable radial flow  \cite{Bozek:2012fw}.

\subsection{Nuclear modification factors ($R_{AA}$ and $R_{CP}$)}

The Nuclear Modification Factor ($R_{AA}$) is a tool for investigating parton-medium interaction mechanisms, without being influenced by the medium's evolution.
Figure \ref{figure3} represents the $R_{AA}$ of $D^{0}$ and $D^{\pm}$ mesons for different centrality bins: 0-10\%, 10-40\%, and 40-80\%, with reference to the $D^{0}$ meson yield in p + p collisions \cite{STAR:2014wif}. To estimate $R_{AA}$, the heavy hadron yield in Au + Au collision is simulated using the HYDJET ++ model, while the heavy hadron yield in p + p collision is taken from the STAR experiment \cite{STAR:2014wif} due to limitations of the model in simulating p + p collision \cite{Lokhtin:2008xi}. We observe that the $R_{AA}$ results of the $D^{0}$ meson from the HYDJET++ model follow the same trend as the experimental data and are consistent with the experimental results \cite{STAR:2018zdy}.

\begin{figure*}
\includegraphics[width=15cm]{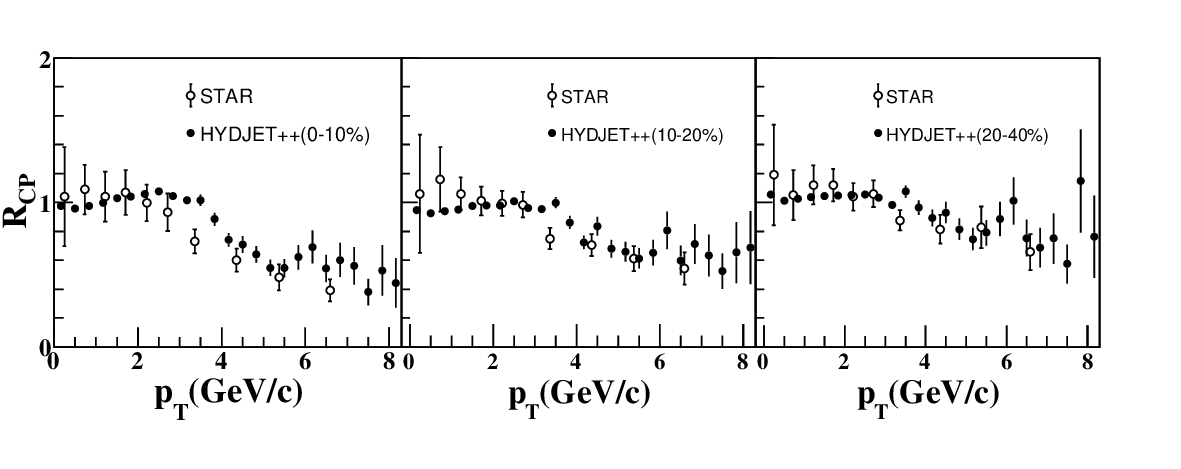}
\caption{\label{figure4} $R_{CP}$ of $D^{0}$ meson as a function of $p_{T}$, using the 40-60\% spectrum as the reference, for the 0-10\%, 10-20\%, and 20-40\% centrality classes in Au + Au collisions. Open markers represent data from STAR measurements \cite{STAR:2018zdy}, while closed markers indicate results from the HYDJET++ model.}  
\end{figure*}

In all centrality bins, a small suppression exists at $p_{T} < 1~\text{GeV/c}$. At low $p_{T}$, the coalescence mechanism of hadronization dominates, which enhances the production of $D_{s}$ and $\Lambda_{c}$ within the QGP, leading to a decrease in the $D^{0}$ production yield in Au + Au collisions. In this $p_{T}$ range,  collisional energy loss dominates over radiative loss. As the momentum of heavy quarks increases, collisional energy loss decreases while radiative loss increases. Thus, at high $p_{T}$, the dominant contribution to energy loss shifts from collisional to radiative loss \cite{Hong:2023cwl, Hong:2025dfj}.

For central (0-10\%) collisions, there is significant suppression in the calculated $R_{AA}$ at $p_{T} > 2.5~\text{GeV/c}$. This suppression decreases as we move from central to peripheral collisions. This trend is expected due to the smaller size of QGP in peripheral collisions compared to central collisions. In peripheral collisions, the $R_{AA}$ value approaches unity, while in most central collisions, the suppression level is about 0.5 at $p_{T} > 3~\text{GeV/c}$. This suppression is attributed to radiative energy loss. At higher $p_{T}$, the suppression is comparable to that light hadrons \cite{STAR:2006uve, PHENIX:2003djd} and electron decays from heavy-flavor hadrons \cite{STAR:2006btx, PHENIX:2005nhb}. Across all centrality, a peak appears when moving from low to intermediate $p_{T}$, indicating the collective motion of charm quarks as they traverse the QGP medium \cite{He:2011qa}.

Furthermore, we compare our results with TAMU \cite{He:2011qa}, SUBATECH \cite{Gossiaux:2010yx}, Torino \cite{Alberico:2011zy}, and Duke \cite{Cao:2013ita} models for the most central collisions. The HYDJET++ model results match well with these models across the entire $p_{T}$ range, except at low $p_{T}$ in the case of the Torino model, due to its limitation in incorporating the coalescence mechanism. The overestimation at low $p_{T}$ by the Torino model is attributed to the lack of an explicit jet quenching mechanism and relies on viscous hydrodynamics to model the QGP medium. Although the HYDJET++ model also omits the coalescence mechanism, but it employs the PYQUEN energy loss model for jet quenching utilizing a boost-invariant ideal hydrodynamic framework, resulting in a more accurate description of $R_{AA}$ at low $p_{T}$.

We also represent the $R_{AA}$ of the $D^{\pm}$ meson for the same three centrality intervals in figure \ref{figure3}. These results are compared with the preliminary data from the STAR experiment \cite{Vanek:2022ekr}. The simulated model results are underestimated but lie within the statistical uncertainty of the experimental data. This discrepancy arises because STAR uses the p + p reference of $D^{*}$ production in the $R_{AA}$ calculation \cite{STAR:2012nbd}, whereas we have used the p + p reference of $D^{0}$ production, as employed in the $R_{AA}$ calculation of the $D^{0}$ meson \cite{STAR:2014wif}. We observed that our model predicts the $R_{AA}$ of $D^{\pm}$ meson at low-$p_{T}$ for the most-central collision. It effectively describes the peak in the intermediate $p_{T}$ region, indicating the collective behavior of quarks within the QGP medium. The $D^{\pm}$ meson exhibits similar suppression  as the $D^{0}$ meson at high $p_{T}$. This suppression decreases as we move from the most central to mid-central and peripheral collisions. At low $p_{T}$, the $D^{\pm}$ meson being more suppressed than the $D^{0}$ meson.

Since the $p_{T}$-spectra for p + p collisions were unavailable for the estimation of $R_{AA}$, as the model is optimized for describing symmetric heavy-ion (AA) collisions with $A\ge 40$ at high energies \cite{Kurt:2014mca}. Hence, the better study of the suppression effects can be performed through the nuclear modification factor ($R_{CP}$). This provides insights into medium-induced modifications without relying on p + p reference data.
Figure \ref{figure4} presents the $R_{CP}$ of $D^{0}$ mesons obtained using the HYDJET++ model for different centrality bins: 0-10\%, 10-20\%, and 20-40\%, with reference to the 40-60\% peripheral centrality bin in Au + Au collisions at $\sqrt{s_{NN}} = 200~\text{GeV}$.
At low $p_{T}$ ($p_{T} < 1~\text{GeV/c}$), the $R_{CP}$ of $D^{0}$ mesons approaches unity. However, at high $p_{T}$, there is significant suppression in the $R_{CP}$ value. This suppression decreases as we move from the most-central to mid-central and peripheral collisions. At high $p_{T}$, radiative energy loss dominates, leading to the observed large suppression in $R_{CP}$. The level of suppression in $R_{CP}$ at high $p_{T}$ is similar to that observed in light hadrons \cite{STAR:2006uve}. However, for light hadrons, a small suppression from unity is also observed at low $p_{T}$ \cite{STAR:2008bgi}, which can be attributed to collective effects such as radial flow, shadowing due to nuclear parton distribution functions, and the dominance of soft physics processes in central collisions. We have also shown results from the STAR experiment \cite{STAR:2018zdy}. We observe that the HYDJET++ model results match well with the experimental data \cite{STAR:2018zdy} for all centrality bins.

\subsection{Antiparticle-to-particle ratio}

\begin{figure*}
\includegraphics[width=15cm]{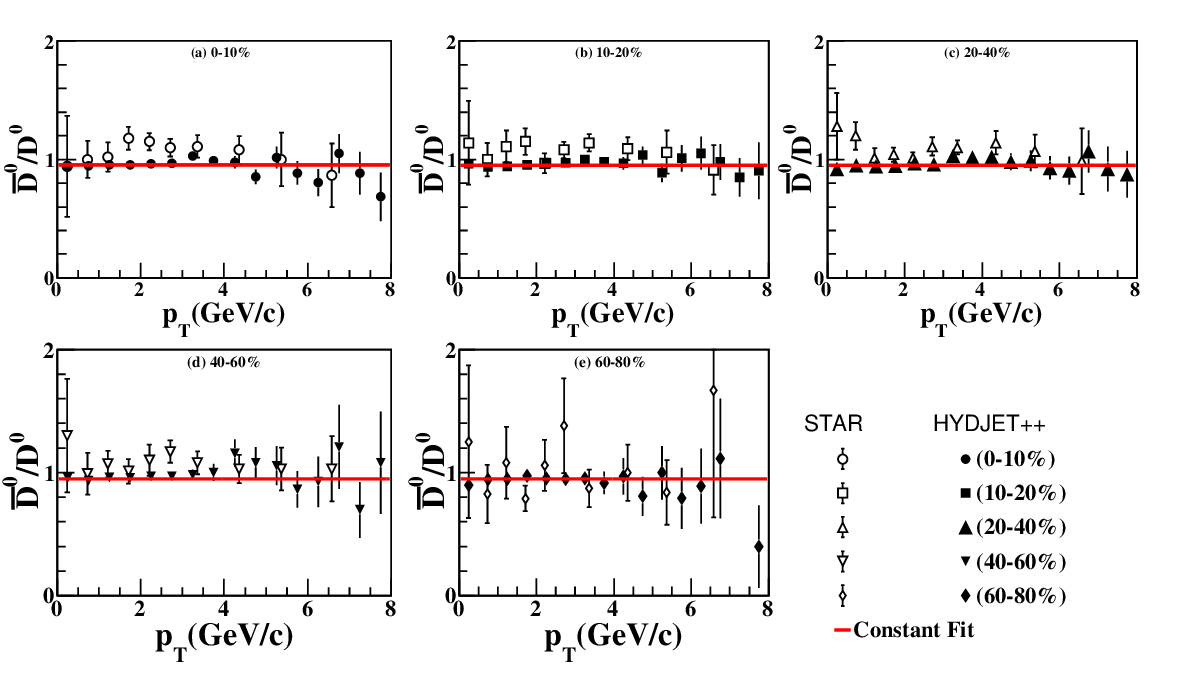}
\caption{\label{figure5} $\overline{D^{0}}/{D^{0}}$ invariant yield as a function of $p_{T}$ at mid-rapidity $(\lvert y \rvert < 1)$ for 0-10\%, 10-20\%, 20-40\%, 40-60\%, and 60-80\% centrality classes in Au + Au collisions at $\sqrt{s_{NN}} = 200$ GeV. Open markers represent data from STAR measurements \cite{STAR:2018zdy}, while closed markers indicate HYDJET++ model results. The red line depicts the constant function that fits the $\overline{D^{0}}/{D^{0}}$ ratios.}    
\end{figure*}

\begin{figure*}
\includegraphics[width=15cm]{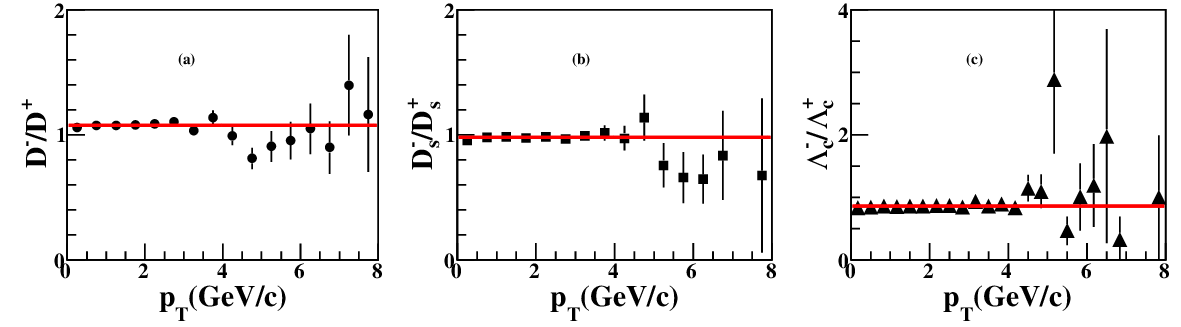}
\caption{\label{figure6} (a) $D^{-}/{D^{+}}$, (b) $D_{s}^{-}/{D_{s}^{+}}$, and (c) $\Lambda_{c}^{-}/\Lambda_{c}^{+}$ invariant yield as a function of $p_{T}$ at mid-rapidity $(\lvert y \rvert < 1)$ for the 0-10\% centrality class in Au + Au collisions at $\sqrt{s_{NN}} = 200$ GeV, as predicted by the HYDJET++ model. The red line depicts the constant function fit.}  
\end{figure*}

In figure \ref{figure5}, the $\overline{D^{0}}/D^{0}$ meson ratio is shown for different centrality bins: 0-10\%, 10-20\%, 20-40\%, 40-60\%, and 60-80\%. The solid line represents a constant function fit. The fitted values of the $\overline{D^{0}}/D^{0}$ invariant yield, obtained from the HYDJET++ model and the STAR experiment \cite{STAR:2018zdy}, are listed in table \ref{table2}. The model results follow a trend similar to the experimental data and show a minimal deviation from unity in central and mid-central collisions. Since charm and anti-charm quarks are produced in pairs, their total amount should be the same. To further explore the charm and anti-charm conservation, it is also essential to calculate the anti-baryon-to-baryon ratios.

\begin{table}[tbp]
    \centering
    \begin{tabular}{|c|c| c|}
    \hline    Centrality &${\overline{D^{0}}}/{D^{0}}$ (HYDJET++)& ${\overline{D^{0}}}/{D^{0}}$ (STAR)\\
    \hline \hline
     0-10\%    & 0.953 & 1.104\\
    10-20\%     & 0.949 & 1.071\\
     20-40\%    & 0.949 & 1.060\\
     40-60\%    & 0.949 & 1.073\\
     60-80\%    & 0.949 & 0.943 \\
     \hline
    \end{tabular}
    \caption{\label{table2}${\overline{D^{0}}}/{D^{0}}$ invariant yield from the HYDJET++ model and STAR results \cite{STAR:2018zdy}, fitted with a constant function for various centrality bins.}
    
\end{table}
Hence, the $D^{-}/D^{+}$, $D_{s}^{-}/D_{s}^{+}$, and $\Lambda_{c}^{-}/\Lambda_{c}^{+}$ ratios as functions of transverse momentum at mid-rapidity $(\lvert y \rvert < 1)$ are shown in figure \ref{figure6} for the most central Au + Au collisions at $\sqrt{s_{NN}} = 200~\text{GeV}$. The constant fit value for $D^{-}/D^{+}$ is 1.076. This is attributed to the charm chemical potential ($\mu_{c}$) of -0.004 GeV, suggesting that anti-charm mesons are produced in slightly greater numbers than charm mesons. The $D_{s}^{-}/D_{s}^{+}$ ratio has a constant fit value of 0.979, which is due to the strange chemical potential ($\mu_{s}$) of 0.007 GeV. This indicates strangeness enhancement within the QGP, which is one of the key signatures of the QGP medium \cite{Singh:1992sp}. The constant fit value for $\Lambda_{c}^{-}/\Lambda_{c}^{+}$ is 0.856. This results from the baryon chemical potential ($\mu_{B}$) of 0.0285 GeV, indicating a baryon-rich region, which is expected to exist at RHIC energies \cite{Lee:2007wr}. 

\begin{figure*}
\includegraphics[width=14cm]{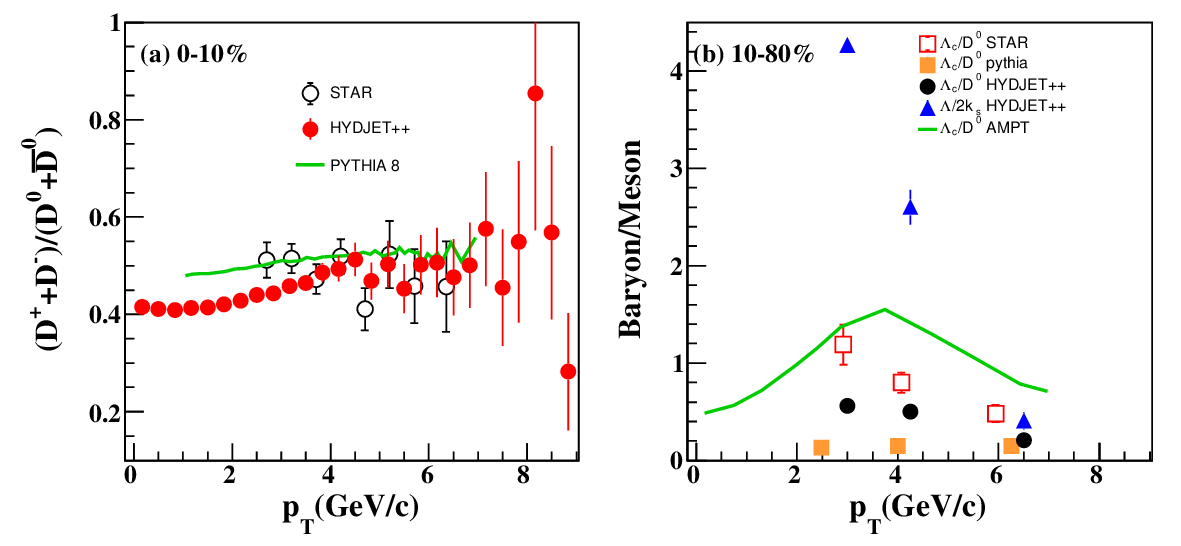}
\caption{\label{figure7} (a) $D^{\pm}/D^{0}$ ratio for the 0-10\% centrality as a function of $p_{T}$ at mid-rapidity $(\lvert y \rvert < 1)$ in Au + Au collisions at $\sqrt{s_{NN}} = 200$ GeV obtained from the HYDJET++ model and compared with data from STAR measurements \cite{Vanek:2022ekr}. Open markers represent data from STAR measurements, while closed markers indicate results from the HYDJET++ model. The green line shows the PYTHIA result \cite{Sjostrand:2014zea}. (b) The baryon-to-meson ratio for the 10-80\% centrality class as a function of $p_{T}$ at mid-rapidity $(\lvert y \rvert < 1)$ in Au + Au collisions at $\sqrt{s_{NN}} = 200$ GeV, obtained from the HYDJET++ model and compared with data from STAR measurements \cite{Radhakrishnan:2019gbl}. Blue and black markers show the $\Lambda/2K_{s}$ and $\Lambda_{c}/D^{0}$ ratio from the HYDJET++ model, while orange markers show the results from Pythia \cite{Sjostrand:2006za}. The green line shows the AMPT model results \cite{Lin:2021mdn}.}  
\end{figure*}

\subsection{Mixed particle ratios}

The left panel of figure \ref{figure7} shows the $D^{\pm}/D^{0}$ yield ratio in the 0–10\% centrality bin. Our model matches well with the experimental data and PYTHIA 8 calculations \cite{Sjostrand:2014zea} at high $p_{T}$. This indicates that the ratio is unaffected by Au + Au collisions compared to p + p collisions. This behavior suggests that both mesons follow the same suppression and hadronization mechanisms in Au + Au collisions.

The right panel of figure \ref{figure7} shows the baryon-to-meson ratio for the 10-80\% centrality bin in Au + Au collisions for both heavy and strange hadrons. The simulated model results for heavy hadrons exhibit the same $p_{T}$ dependence (a decreasing trend) as strange hadrons. Similarly, the STAR experimental data also show a decreasing trend with $p_{T}$ \cite{Radhakrishnan:2019gbl}. The baryon-to-meson yield ratio for all hadrons is enhanced compared to PYTHIA predictions for p + p collisions \cite{Sjostrand:2006za}. This enhancement is attributed to quark coalescence within the QGP medium for both light and heavy hadrons. However, the HYDJET++ results for heavy hadrons show a significant deviation from the STAR results \cite{Radhakrishnan:2019gbl}, which is due to the absence of the quark coalescence mechanism in the model. We have also compared our simulated results with the improved AMPT model results \cite{Lin:2021mdn}. Our model results underpredict the AMPT estimation, but the AMPT results also overpredict the experimental data. This discrepancy is attributed to the lack of radiative energy loss in the AMPT model.

The distinct nature of the $\Lambda_{c}$ baryon, compared to the $D^{0}$ and $D^{\pm}$ mesons, indicates that nuclear effects in Au + Au collisions have little impact on charm quark production. However, the medium modifies the hadronization process, leading to a redistribution of charm quarks among different open charm hadron species.

\section{Summary and Outlook}
\label{sec:summ}
In this study, we have investigated the production of heavy hadrons, including $D^{0}$, $\overline{D}^{0}$, $\Lambda_{c}$, $D^{+}$, and $D^{-}$, in Au + Au collisions at $\sqrt{s_{NN}} = 200$ GeV for different centrality bins using the HYDJET++ model. This analysis reveals decreasing trend of $\gamma_{c}$ from central to peripheral collisions, reflecting reduced system size in peripheral collision. Additionally, $\gamma_{c}$ shows mass dependence, with higher values for heavier charm hadrons in central collisions. Moreover, the $p_{T}$ spectra results demonstrate that the model describes the experimental data reasonably at low and intermediate $p_{T}$, capturing the essential features of charm hadron production in a QGP medium. At high $p_{T}$, the model overpredicts the experimental data across all centrality bins, indicating that further refinements are needed to improve the description of charm quark energy loss. The $p_{T}$ spectra further indicate that central collisions produce a hotter and denser QGP compared to peripheral collisions.  Furthermore, $p_{T}$ spectra indicate that heavy hadrons thermalized earlier than light hadrons due to their larger mass. The nuclear modification factors ($R_{AA}$ and $R_{CP}$) as a function of $p_{T}$ for $D^{0}$ mesons are in good agreement with experimental measurements, particularly at low and intermediate $p_{T}$. In the $R_{AA}$ and $R_{CP}$ results, significant suppression is observed in central collisions compared to peripheral collisions, which is consistent with experimental findings. The suppression at high $p_{T}$ confirms the dominance of radiative energy loss mechanisms, while the suppression at low $p_{T}$ is attributed to collective effects, such as coalescence and radial flow. Comparisons of HYDJET++ with other theoretical models, such as TAMU, SUBATECH, and Duke, show good agreement in most cases, particularly at low and intermediate $p_{T}$. However, the Torino model exhibits discrepancies at low $p_{T}$. This is because it omits the coalescence mechanism and lacks an explicit jet quenching model. The study of antiparticle-to-particle ratios demonstrates that the model results are consistent with experimental data, reaffirming the chemical potential within the QGP medium. Additionally, the $D^{\pm}/D^{0}$ yield ratio in most central collisions aligns well with experimental data and PYTHIA 8 at high $p_{T}$, indicating similar suppression and hadronization processes in both Au + Au and p + p collisions. The baryon-to-meson ratio for heavy hadrons follows a similar trend as observed for light hadrons and experimental data, but underpredicts the observed values due to the absence of heavy quark coalescence with in-medium light quarks in the HYDJET++ model. This underprediction highlights the importance of incorporating coalescence mechanisms into theoretical models to improve their ability to describe baryon production in the QGP. 

Overall, this study demonstrates the capability of the HYDJET++ model to describe the production of charm hadrons, centrality dependence, $R_{AA}$ and $R_{CP}$ behavior, antiparticle-to-particle ratios, and baryon-to-meson ratios in heavy-ion collisions. However, the overprediction at high $p_{T}$ indicates the need for further improvements in modeling energy loss mechanisms for heavy quarks. These findings motivate future investigations into other QGP signals, including elliptic flow, which will be the focus of our next work. This study also encourages further exploration of heavy quarkonia, such as $J/\psi$, $\psi^{'}$, and $\chi_{c}$, by readjusting the charm enhancement factor ($\gamma_{c}$) in the HYDJET++ model at RHIC and LHC energies. The study of quarkonia production might provide deeper knowledge of the dynamics of heavy quarks in the QGP medium and enhance our understanding of jet quenching and other in-medium effects.

\begin{acknowledgments}
AK sincerely acknowledges financial support from the Institution of Eminence
(IoE), BHU. SS and RG acknowledge the financial support from UGC under the
research fellowship scheme at central universities.
\end{acknowledgments}

\bibliography{apssamp}

\end{document}